# Exciton and Carrier Dynamics in 2D Perovskites

Andrés Burgos-Caminal, Etienne Socie, Marine E. F. Bouduban, and Jacques-E. Moser *

Photochemical Dynamics Group, Institute of Chemical Sciences and Engineering,
and Lausanne Centre for Ultrafast Science (LACUS),
École Polytechnique Fédérale de Lausanne, 1015 Lausanne, Switzerland

* Corresponding author, email: je.moser@epfl.ch

## Abstract

Two-dimensional Ruddlesden-Popper hybrid lead halide perovskites have become a major topic in perovskite optoelectronics. Here, we aim to unravel the ultrafast dynamics governing the evolution of charge carriers and excitons in these materials. Using a combination of ultrabroadband time-resolved THz (TRTS) and fluorescence upconversion spectroscopies, we find that sequential carrier cooling and exciton formation best explain the observed dynamics, where exciton-exciton interactions play an important role in the form of Auger heating and biexciton formation. We show that the presence of a longer-lived population of carriers is due to these processes and not to a Mott transition. Therefore, excitons still dominate at laser excitation densities. We use kinetic modeling to compare the phenethylammonium and butylammonium organic cations while investigating the stability of the resulting films. In addition, we demonstrate the capability of using ultrabroadband TRTS to study excitons in large binding energy semiconductors through spectral analysis at room temperature.

## TOC Graphic

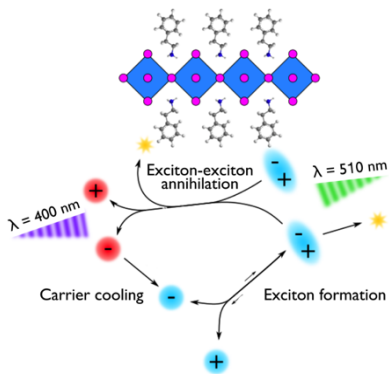

Lead halide perovskites (LHPs) have become a widely studied and applied family of semiconductors in the field of optoelectronics, especially towards the development of solar cells that can compete with well-established technologies. The current solar-to-energy conversion has been pushed beyond 25 %[1] owing to the quick and continued improvements.[2–5] One of the key limitations of this technology has been its stability with respect to ambient conditions such as temperature, light and, especially, humidity. To tackle this problem, one of the proposed solutions is to use 2D perovskites such as those of the Ruddlesden-Popper family to increase the stability with respect to moisture.[6,7] Mixing compositions of 2D/3D LHPs has been shown to increase the stability while maintaining a good performance.[8,9] These lower dimensionality perovskites are obtained by introducing larger cations into the precursor mix. In the perovskite structure, characterized by the formula $ABX_3$, the A cation is a small cation that can fit in the center of the cube formed by every eight $BX_6$ octahedra (in this case, B is $Pb^{2+}$ and X is $I^-$). If the A cation is replaced by a large cation, such as phenethylammonium ($C_6H_5C_2H_4NH_3^+$ or PEA), then the 3D crystalline structure can no longer be formed. Instead, a layered structure consisting of $PbI_6$ octahedra on the same plane surrounded by the large cations is formed. This effectively produces a quantum confinement effect due to the thickness of the layer that greatly affects the electronic properties of the material. Variable stoichiometric proportions of small cations can be added to produce multilayered quasi-2D perovskites. These can be described as quantum wells of quantized thickness, depending on the number of layers. A general formula of $R_2A_{n-1}B_nX_{3n+1}$ can be defined for Ruddlesden-Popper 2D perovskites, where R is the large monovalent cation and n is the number of layers. However, it is of great difficulty to obtain phase-pure samples of multilayered 2D perovskites. Efforts towards this goal have been made for single crystals[10] and thin films.[11]

In such low-dimensionality systems, there is not only quantum confinement but also dielectric confinement due to the difference in the permittivity between the ionic perovskite layer and the bulky organic cation. These two effects enhance electron-hole correlations, increasing the binding energy ($E_b$) of excitons from tens of meV in 3D $CH_3NH_3PbI_3$ to hundreds of meV. Furthermore, the variations in $E_b$ with the cation can be attributed to the dielectric confinement due to the varying permittivity.[12–14] Given the considerable $E_b$ in 2D perovskites, it is safe to assume that photoexcitations will have a strong excitonic character. Nonetheless, it is possible to obtain solar cells for at least the n = 3 composition.[6]

Charge carriers excited with excess energy are known as hot carriers and follow a characteristic evolution of their relaxation that varies with the material. Typically, upon photoexcitation, charge carriers are generated in a distribution that has to undergo a thermalization process before being described by Fermi-Dirac statistics as a thermal distribution. This process can take less than 85 fs in 3D perovskites.[15] The charge carrier distribution is then characterized by a carrier temperature ($T_c$), which is higher than that of the lattice ($T_l$). Therefore, it subsequently cools down through phonon-carrier interactions until both temperatures converge. This initially occurs through optical phonon emission until the temperature reaches the energy of said phonons and, if necessary, then continues with the involvement of acoustic phonons. In LHPs, the longitudinal optical (LO) phonon energy lies below the room temperature energy. Hence, the involvement of acoustic phonons is negligible.[16] When comparing hot carrier cooling between 2D and bulk perovskites, it has been found that the process occurs considerably faster in the confined perovskite, at least in colloidal suspensions.[17] This was explained by taking into account the decreased Coulomb screening by the lattice due to the dielectric confinement. Indeed, the dielectric constant of the organic cation was recently found to significantly affect the cooling process.[18] On the other hand, carrier cooling was found to be lengthened in perovskite nanoparticles, allowing the extraction of hot carriers.[19]

There is still uncertainty regarding the process of hot carrier generation and cooling and exciton formation, as well as whether hot excitons are formed. A direct probe of excitons is extremely useful to completely reveal the early dynamics in these systems. Two techniques that have the potential to selectively probe excitons are time-resolved photoluminescence through fluorescence upconversion spectroscopy (FLUPS) and time-resolved THz spectroscopy (TRTS). While the latter is typically used to selectively study charge carriers, it can also be used to identify excitons. This is achieved through differentiated spectral responses of the two species.[20–25] However, the studies carried out in the past on exciton formation using TRTS have been limited to the 0.5-3 THz range. This requires an $E_b$ close to this range to observe distinct signatures, which demands the use of cryogenic temperatures to stabilize the excitons. Here, we want to not only study the exciton-carrier dynamics of 2D perovskites but also demonstrate the capability of using an ultrabroadband TRTS setup (up to 20 THz) to study excitons at room temperature. We combine this technique with FLUPS to further support our findings and observe the effect produced by different cation compositions.

To study the carrier-exciton dynamics in 2D perovskites, we need samples that are transparent to our broadband THz pulses in our TRTS setup. We used thin films to facilitate spectroscopy measurements in transmission mode and to be close to the device operating conditions of the material. As previously reported,[26] this can be achieved using high-density polyethylene (HDPE) as a substrate. These samples were used for both TRTS and FLUPS to minimize the variations. More information about the samples can be found in the SI (Section S5) and the Methods section.

TRTS is especially sensitive to charge carriers. The THz absorption signal ($-\Delta E/E$) is, for a thin film, typically proportional to the photoconductivity ($\Delta\sigma$), which in turn is proportional to the photogenerated carrier density ($N$) and mobility ($\mu$), following

$$\frac{-\Delta E}{E} \propto \Delta\sigma = \mu \cdot e \cdot N \tag{1}$$

This can be used to rationalize the frequency-averaged THz absorption dynamics (measured at the point of highest THz electric field). Nonetheless, this is only valid if the absorption spectrum is flat across different frequencies and does not substantially change over time, such as in the case of exciton formation. Otherwise, it is important to obtain spectra at different time delays to better characterize the system. This will be explored later in this article. On the other hand, the fluorescence can be considered proportional to the exciton density. This is due to the large $E_b$ that induces exciton formation upon carrier relaxation before recombination. Since each exciton has a certain probability of recombining and emitting a photon, according to the radiative recombination kinetic constant, the higher the density is, the more photons will be emitted at a given time. Thus, each technique has a different sensitivity. While the FLUPS signal will be dominated by excitons, free charge carriers will be more important in TRTS. This can be seen in Figure 1, where the THz signal decreases as the fluorescence increases. This clearly indicates a depletion of charge carriers through exciton formation. This process occurs during the first ps after excitation until a quasi-equilibrium is reached. After that, exciton recombination dominates the dynamics. The fit is the result of a tentative triexponential global fit where the time constants are linked between the two experiments, resulting in $\tau_1$ = 0.35 ps, $\tau_2$ = 7 ps and $\tau_3$ = 60 ps. A full view of the decays is shown in the SI (Figure S1). *A priori*, $\tau_1$ can be assigned to carrier relaxation and exciton formation, while the other two constants describe a combination of bimolecular and monomolecular decays highly dependent on the pump fluence. More evidence can be found when comparing the TRTS decays for samples excited at 400 nm and 510 nm (Figure 1, right). In the former, a higher carrier population is formed that needs to relax and condense to reach equilibrium. This does not occur in the latter since a large population of excitons is formed from the beginning. Figure S2 shows how the decays are very similar after the initial exciton formation decay.

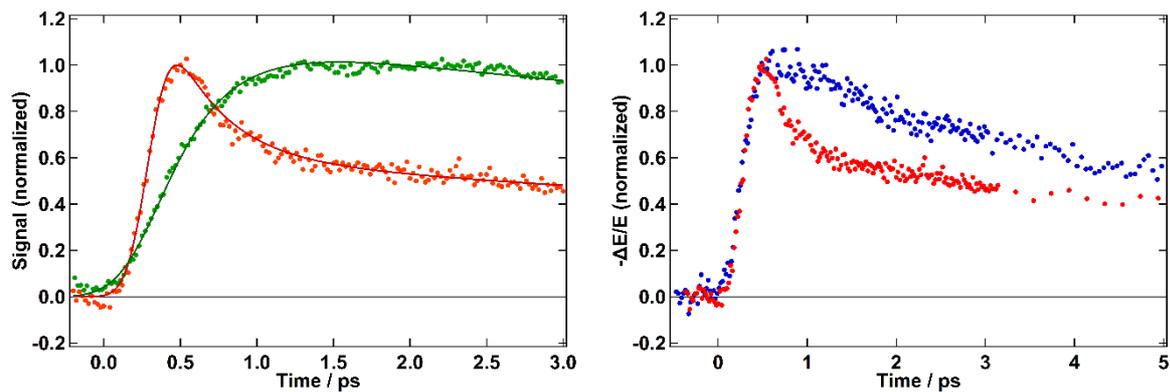

**Figure 1**. Left: Simultaneous rise and decay of the FLUPS (green, 55 µJ cm$^{-2}$, $\lambda_{obs}$ = 532 - 538 nm) and TRTS (red, 56 µJ cm$^{-2}$) signals at $\lambda_{ex}$ = 400 nm. Right: TRTS decays at $\lambda_{ex}$ = 400 nm (red, 56 µJ cm$^{-2}$) and $\lambda_{ex}$ = 510 nm (blue, 60 µJ cm$^{-2}$). The TRTS signal is obtained at the maximum point (see Figure 2E).

Let us now focus on the TRTS evolution using the spectral response. The fact that there is a considerable signal rise with the same time resolution for both excitation energies indicates a noticeable THz absorption coming from the excitons. Otherwise, in the 510 nm case, a certain delay would be required to build up some carrier population. Equally, from a simple thermodynamic point of view, an equilibrium of mostly excitons should be attained after several ps when exciting above the bandgap into the continuum. In fact, we can use the Saha equation to estimate the expected proportion of carriers.[27]

$$\frac{x^2}{1-x} = \frac{1}{n} \cdot \frac{2\pi \cdot \mu \cdot k_B T}{h^2} \cdot \exp\left(\frac{E_b}{k_B T}\right) \quad , \tag{2}$$

where $\mu$ is the reduced mass of the exciton (0.28 $m_0$)[28], $E_b$ is the exciton binding energy, $k_B$ is the Boltzmann constant, $T$ is the temperature, h is the Planck constant, $n$ is the excitation density and $x$ is the fraction of free carriers. At $T$ = 300 K, considering an $E_b$ of 270 meV [28] with an excitation density of $10^{18}$ cm$^{-3}$, the expected carrier fraction is less than 1%. Our experiments can reach excitation densities of $10^{19}$ cm$^{-3}$ for the highest fluences, further decreasing the free charge carrier fraction. Therefore, the bulk of the signal when excited at 510 nm and at later times should come from excitons according to this analysis. In contrast, at high excitation densities, semiconductors undergo an insulator-to-metal transition or Mott transition, where the screening produced by the carriers effectively reduces the exciton binding energy until the attraction force is lost. The result is an electron-hole plasma. This can be observed with TRTS of Si.[29] Recently, it was found that excitons are surprisingly persistent in bulk CH$_3$NH$_3$PbBr$_3$ perovskite above the Mott density, which is attributed to Mahan excitons.[30] On the other hand, from a layered perspective, F. Thouin *et al.* estimated the Mott density at 2.2·10$^{11}$ excitons cm$^{-2}$ from a Bohr radius of 12 nm. Meanwhile, from the penetration depth and the layer thickness we can calculate an average exciton density per layer of 6.7·10$^{11}$ excitons cm$^{-2}$, for our highest fluence (120 μJ cm$^{-2}$ at $\lambda_{ex}$ = 400 nm).[31] Although the Mott density may be underestimated since neither dielectric confinement nor exciton renormalization effects were taken into account, as stated by the authors, it is unclear whether an electron-hole plasma is stabilized at these densities. This motivates an analysis of the THz absorption spectra at different time delays to discern the evolution of free carriers into excitons. Both species are expected to show characteristic THz spectra. Charge carriers are typically described using the semiclassical Drude and Drude-Smith models, where the latter is used to describe carriers with an important backscattering component, usually observed in quantum confined systems such as nanoparticles:[32,33]

$$\Delta\sigma(\omega) = \frac{\omega_p^2 \cdot \varepsilon_0 \cdot \tau}{1 - i\omega\tau} \quad , \tag{3}$$

$$\Delta\sigma(\omega) = \frac{\omega_p^2 \cdot \varepsilon_0 \cdot \tau}{1 - i\omega\tau} \cdot \left(1 + \frac{c}{1 - i\omega\tau}\right) \quad , \tag{4}$$

where $\omega_p = \sqrt{\dfrac{Ne^2}{\varepsilon_0 m^*}}$ is the plasma frequency, $N$ is the charge carrier density, $m^*$ is the effective mass, $\tau$ is the scattering rate, and $c$ is the Drude-Smith backscattering coefficient with values ranging from $c = 0$ (pure Drude) to $c = -1$. Both models are expressed as a function of angular frequency ($\omega$). These models consider that charges are accelerated under the THz electric field and are scattered with a time constant $\tau$.[34] This acceleration produces THz absorption, and the result is a frequency-dependent photoconductivity $\Delta\sigma(\omega)$, where $\Delta\sigma(0)$ is defined as the DC photoconductivity or $\Delta\sigma_{DC}$. It has the values $\Delta\sigma_{DC} = \omega_p^2 \varepsilon_0 \tau$ and $\Delta\sigma_{DC} = \omega_p^2 \varepsilon_0 \tau (1 + c)$ for the Drude and Drude-Smith models, respectively. Therefore, $\Delta\sigma_{DC}$ typically has a nonzero value for charge carriers unless $c = -1$. In contrast, excitons do not follow these models in the interaction with THz radiation. In this case, the THz absorption corresponds to intraexcitonic transitions between internal degrees of freedom (1s-np transitions) and between the excitonic and free carrier states (Scheme 1).[20,35] Such transitions have been studied in the past, *e.g.*, in GaAs quantum wells[20-22] or bulk lead halide perovskites,[35] at low temperatures. Accurate modeling needs to take into account multiple transitions.[20] However, considering the high binding energy in our sample, the different intraexcitonic transitions can be separated and approximated as a Lorentz oscillator.[35] In addition, they are expected to peak outside the observed window. The model follows the equation:[33]

$$\Delta\sigma(\omega) = -\dfrac{A\,\omega\,\mathrm{i}}{\omega_0^2 - \omega^2 - \mathrm{i}\gamma\omega}, \qquad (5)$$

where $A$ is the amplitude, $\omega_0$ is the resonance frequency and $\gamma$ is the width of the resonance. The Lorentz oscillator model is characterized by a zero $\Delta\sigma_{DC}$. Hence, any $\Delta\sigma_{DC} \neq 0$ can be assigned to the free carrier contribution. At low frequencies, an excitonic model predicts a greater imaginary part of $\Delta\sigma(\omega)$, seen as a phase shift. This is expressed as a measure of the polarizability of excitons and can be followed through the changes in $\Delta E$ at the zero-crossing point.[36-39] However, such a phase-shift (imaginary part) component can also be produced by out-of-phase motion of carriers, such as the backscattering scenario of nanostructured materials modeled with the Drude-Smith model. This is why a full study of the $\Delta\sigma(\omega)$ spectra at different delays is preferable and more selective.

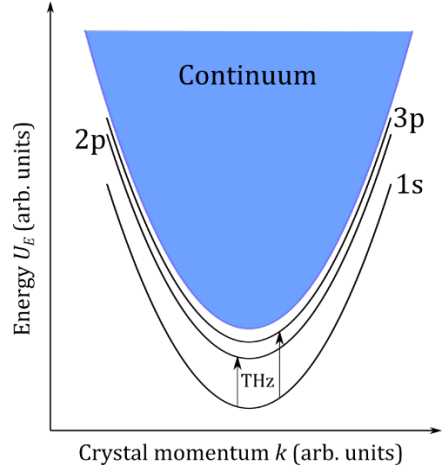

**Scheme 1**. Intraexcitonic transitions accessible with THz spectroscopy.

We performed a 2D TRTS scan of our PEA$_2$PbI$_4$ sample excited at 400 nm with an energy fluence of 60 µJ cm$^{-2}$. The 2D maps are shown in Figure S3. A selection of the progressive spectral changes can be observed in Figure 2A. No clear resonance is observed in the analyzed frequency range. Resonances, such as those observed for GaAs quantum wells,[20] are not expected to be fully observable given the large $E_b$ and the fact that our probe can only reach *ca.* 80 meV. Nonetheless, the response below the peak can still be observed. As the signal evolves, the spectrum changes through a marked decrease in $\Delta\sigma_{DC}$ while Im($\Delta\sigma$) increases and then is maintained at a large value. This can be preliminarily correlated with exciton formation. However, stronger evidence and information are obtained through the analysis using the photoconductivity models previously presented. If we take a linear combination of the Drude-Smith model and a Lorentz oscillator (red and green, respectively, in Figures 2C and 2D), we can apply a global fit to all the spectra. The results (Figure 2C) show how the amplitude of the Drude-Smith model (carriers) decreases with the general decay of the signal, while that of the Lorentz oscillator (excitons) increases and stabilizes. This coincides with what we observed in Figure 1. Certainly, the global fit is a simplification. Changes affecting $\tau$ or $\gamma$ involving many-body effects are not considered. Nevertheless, the fit gives a convincing description of the global process. The effects of the spectral changes can be observed in the $\Delta E(t)$ waveform as a gradual phase shift, which should agree with its measurement using the zero-crossing point. However, $\Delta E$(–0.1 ps) already possesses a considerable signal at the zero crossing, which is probably produced by out-of-phase carriers. The results of the fit show that the excitons are described by a resonance centered at $\omega_0$ = 26 THz with $\gamma$ = 20 THz, while the carriers are described by the Drude-Smith model with $\tau$ = 6.7 fs and $c$ = −0.7. While the values cannot be very precise due to the similarities of the two models, the latter describes carriers with a considerably low mobility ($\mu_{DC}$ = 5.8 cm$^2$ V$^{-1}$ s$^{-1}$, considering an effective mass of 0.605 $m_0$ [28]) and an important backscattering component. This is not surprising, taking into account the moderate mobility of perovskites[40] and the nanostructured nature of the sample.

Interestingly, while the excitonic contribution starts to dominate after a few ps, there is still a noticeable presence of carriers, judging from the value of $\Delta\sigma_{DC}$. Here, we can suggest two hypotheses. On the one hand, Auger heating through exciton-exciton annihilation may produce new hot carriers for every annihilated exciton.[41,42] On the other hand, the high excitation density used may produce a partial Mott transition. In this case, a portion of the excitons are or remain dissociated as an electron-hole plasma. Nonetheless, it is clear from the spectral analysis that a great proportion of the excitons are undissociated.

One way to clarify this observation is by analyzing the emission spectra at different fluences. In Figure S4, we can see the effect of fluence on the emission spectrum at an early time (1 ps). If we were in a regime where the Mott transition was starting to play a role, then $E_b$ should decrease with fluence as the screening becomes stronger.[30] This would translate into a blueshift of the emission peak since the excitons are less stabilized. However, we do not observe any blueshift of the peak with fluence. In contrast, a shoulder on the red side of the peak becomes more important as we increase the fluence. This could be a sign of biexcitons, where the additional stabilization produces a slight redshift of the emission that is observed as a small additional sideband. This was described for colloidal nanoplatelets in a previous work.[43] In addition, biexcitons have already been reported in 2D perovskites with a binding energy of 44±5 meV by Thouin *et al*.[31] Indeed, qualitative agreement can be found through a multipeak fit (Figure S5). The biexciton picture reinforces the idea of important exciton-exciton interactions being the source of new carriers through an Auger process. Interestingly, a shift or broadening towards the blue can be observed during the first 10 ps that is not fluence dependent (Figure S6). A possible explanation is that it may be produced by reabsorption on the blue side of the emission by the excitonic absorption peak. This would be due to the geometry of the measurement, where the emitted light is collected from the back. As the excitons diffuse across the sample less light is reabsorbed and a blue-shift is observed.

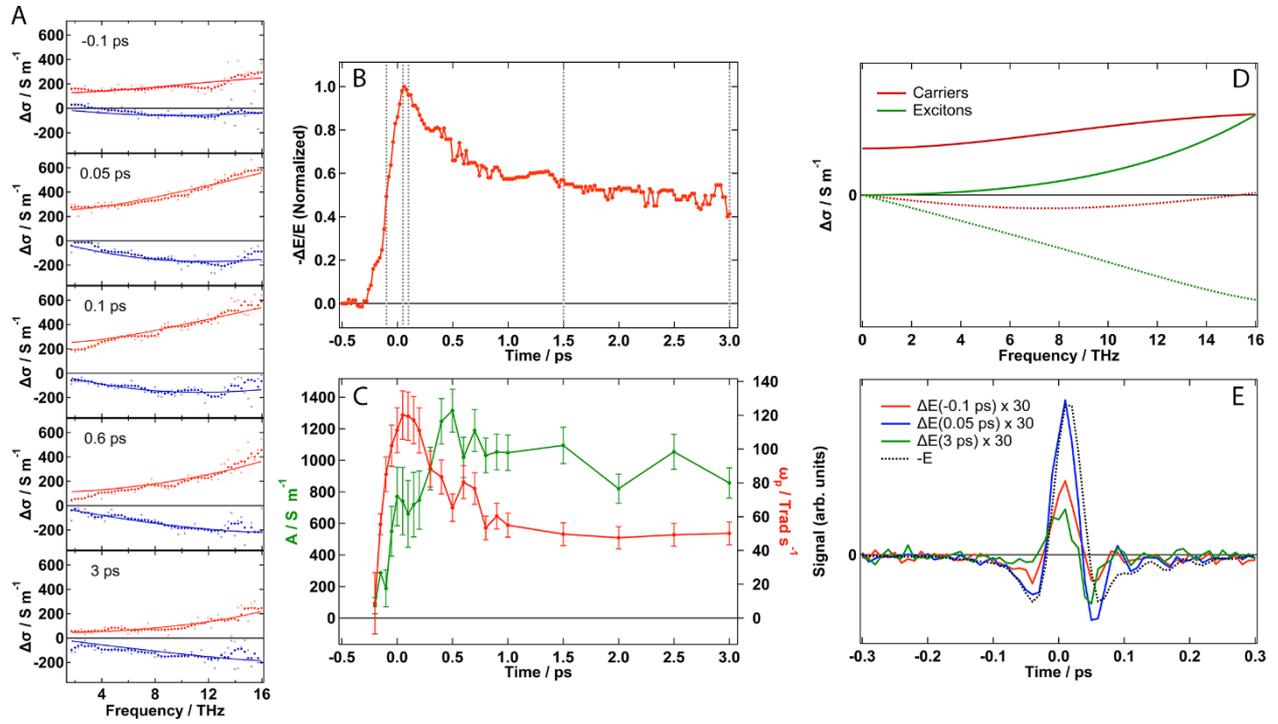

**Figure 2.** A) Real (red) and imaginary (blue) parts of the $\Delta\sigma$ spectra at different time delays after excitation. The solid lines are the results of the global fit. B) $-\Delta E/E$ signal evolution as a function of pump-probe delay. The dashed lines indicate the times corresponding to the spectra in A. C) Result of the global fit of the spectra at all times. $\omega_p$ and $A$ are taken as a measure of the amplitude of the signals emanating from carriers and excitons, respectively, to follow their evolution. D) Simulated spectra from the Drude-Smith (red) and Lorentz oscillator (green) models as a result of the global fit. E) $\Delta E$ (solid) and $-E$ (dashed) THz waveforms. The plot shows the changes in the waveforms that produce the spectral changes shown in A. The measurements are taken at $\lambda_{ex}$ = 400 nm and $F$ = 60 µJ cm$^{-2}$ on PEA$_2$PbI$_4$. In A and B, median filters are applied to clean up the traces.

To gain more insights into the kinetics of the processes involved, it is useful to carry out fluence-dependent measurements, where the higher order kinetics will be highlighted. When exciting with excess energy ($\lambda_{ex}$ = 400 nm, Figure 2C), the increased fluence mainly affects the end of the initial decay and the long-term decay in opposite ways. However, the relative amplitude of the initial decay is surprisingly little affected by a tenfold increase in fluence. If a simple equilibrium between carriers and excitons was established, then it would be expected that the greatest initial decay occurs at higher fluences due to the facilitation of nongeminate exciton formation at higher densities. This is not the case, and the reason may lie in the impossibility of hot exciton formation. In fact, we do not observe hot exciton emission in the way that it was observed for nanoplatelets and nanoparticles.[43,44] This initial decay may therefore be limited by cooling, which can be lengthened at higher fluences due to phonon bottleneck effects[45], Auger heating,[46,47] and polaron screening.[26,48,49] Afterwards, a carrier-exciton equilibrium is formed, and the decay dynamics are dominated by exciton recombination. At higher fluences, bimolecular exciton-exciton annihilation[42] becomes increasingly important, shortening the decay of both carriers and excitons. These two phenomena can be observed with a triexponential fit as a slight increase in the first time constant and a decrease in the other two (Table S1).

Alternatively, when exciting at resonance conditions ($\lambda_{ex}$ = 510 nm, Figure 3D), there are two observable effects of the fluence: an expected decrease in the general decay lifetime and the appearance of a slower rise at the lowest fluence used. This rise could be assigned to the formation of charge carriers, from the initial excitons to when equilibrium is reached. However, taking into account that the exciton binding energy is on the order of 270 meV,[28] much higher than the thermal energy ($k_B T$ = 25.6 meV at 25°C), it is difficult to assume that excitons can spontaneously split into carriers only due to thermal energy. Once again, Auger heating through exciton-exciton annihilation can be used to explain the rise. To do so, we can define three fluence regimes: a) a low fluence regime, where no such processes occur, b) an intermediate fluence regime, where such processes occur and can be observed as a slow buildup of the carrier population, and c) a high fluence regime, where bimolecular decay processes are so important that any rise in the signal is buried. The data we have been able to collect mostly belong to the latter regime, while the lowest fluence we show, 12 µJ cm$^{-2}$, belongs to the intermediate regime.

We can propose a model that includes Auger heating through exciton-exciton annihilation as well as sequential carrier cooling, exciton formation and recombination. The proposed model is described by the scheme in Figure 3A, and the differential equations can be found in the SI. The model considers three different populations, namely, excitons, hot carriers and cold carriers. It considers two possible excitation sources generating either hot carriers ($\lambda_{ex}$ = 400 nm) or excitons ($\lambda_{ex}$ = 510 nm). Excitons can monomolecularly recombine, either through emission of photons or interaction with traps, with a kinetic rate constant $k_{er}$. In addition, they can undergo a process of exciton-exciton annihilation where Auger heating is produced ($k_a$). Two excitons produce two hot carriers from one of them, while the other, having transferred its energy, recombines. As a simplification, hot carriers are monomolecularly cooled with a rate constant $k_c$. Finally, cold carriers condense into excitons ($k_{ef}$), but the latter can also split into cold carriers ($k_{ed}$). However, this process is disfavored due to the large $E_b$. Indeed, several approximations are required to obtain a model we can work with. First, all processes occurring in the first hundreds of fs, defining the rise of the signal, are included in a generation term $G$ with the form of a Gaussian. These processes can include coherence loss,[50] thermalization (not cooling)[15] and the development of many-body interactions,[51] perhaps including polaron formation.[26,48,49,52,53] This explains why the rise is characterized by a Gaussian of FWHM = 300 fs, longer than the time resolution expected for this experiment ( ~ 50 fs). Next, hot and cold carriers are considered as two distinct particles and not as a distribution of particles characterized by a certain temperature. In addition, phonon populations are not considered.[46] Possible phonon bottleneck effects are considered as a change in $k_c$. Finally, processes affecting the actual density of carriers, such as diffusion[54] or even trap saturation, are not included.

Employing numerical methods, we can use this model to simulate the signal obtained from a combination of the densities of the three particles times a proportionality constant (Figure 3B). Clearly, this model can reproduce the two regimes observed at high and moderate fluences as well as be used to extrapolate what the signal would look like at a very low fluence. Moreover, we can use the model to globally fit all the data across the two different wavelengths. The results show general agreement with both the fluence and wavelength trends in Figures 3B and 3C, showing that this three-particle picture is sufficient for explaining the observed behavior. The strength of the global fit lies in the fact that all kinetic parameters are shared for all traces and the density is fixed to a value proportional to the fluence used (see more details in the SI). Certainly, this, in addition to the approximations and possible uncertainties, explains the small deviations. Indeed, perfect agreement could be obtained from independent fits, as for any complex model, but the results would be meaningless. The numerical results are shown in the SI. These should not be taken at face value since different combinations can give acceptable fits, especially for the bimolecular constants. The importance lies in the global behavior and how the model explains the trends. Indeed, the rise at moderate fluence when exciting at $\lambda_{ex}$ = 500 nm can then be explained by the difference in the strength of the signal coming from carriers and excitons (2 cold carriers produce *ca.* 4 times the signal of 1 exciton). Exciton-exciton annihilation ($k_a$) is already dominating the recombination kinetics at this fluence, serving as a source of 2 new carriers for each recombined exciton. Furthermore, the normalized signal contributions from excitons and carriers according to the fitted model (Figure S7) show qualitative agreement with the spectral evolution (Figure 2C).

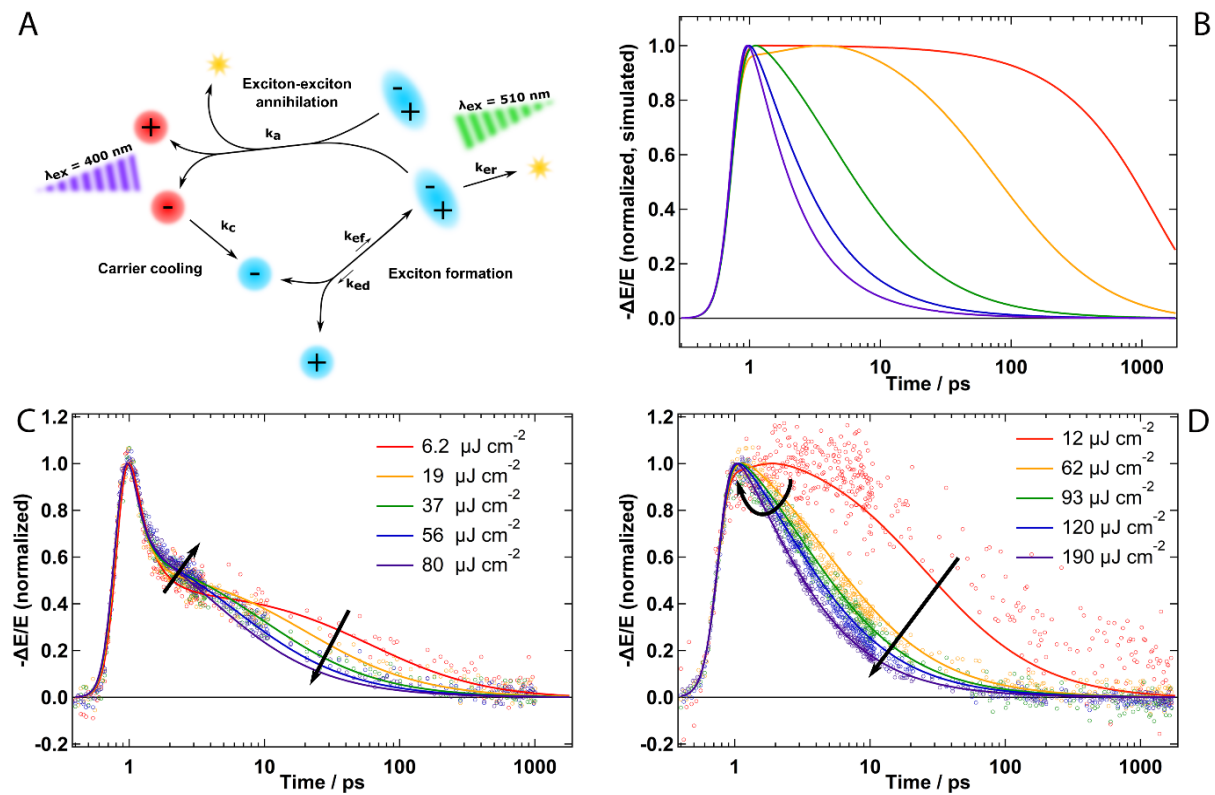

**Figure 3.** A) Kinetic model relating hot carriers (red), cold carriers (blue) and excitons generated with the two wavelengths used in the presented experiments. The kinetic constants describe the processes of carrier cooling ($k_c$), exciton formation ($k_{ef}$), exciton dissociation ($k_{ed}$), exciton recombination ($k_{er}$) and exciton-exciton annihilation ($k_a$). B) Simulations obtained from the model for a wide variety of fluences. 1) At high densities, the signal can decay with strong bimolecular behavior (purple, blue and green). 2) At a low density, it can have a slow exciton recombination decay (red). 3) In the intermediate regime, the signal exhibits a slight rise due to the generation of carriers under Auger heating, followed by recombination (orange). C and D) TRTS data and the results of the global fit for $\lambda_{ex}$ = 400 nm and $\lambda_{ex}$ = 510 nm, respectively. The arrows highlight the effect of the fluence.

Once we have explained the observed signal through both spectral analysis and decay modeling, we can compare the obtained results with those for a different composition. To this aim, we investigated the TRTS dynamics upon $\lambda_{ex}$ = 400 nm excitation of a sample of BuA$_2$PbI$_4$, where BuA stands for butylammonium. Judging from the steady state spectra (Figure S10), the two samples are not very different in terms of $E_b$. Regarding the TRTS dynamics, the main difference is observed in the first decay, which appears to lengthen with increasing fluence (Figure 4). We know that this decay corresponds to the formation of excitons from the initially high population of carriers. In part, this is limited by the cooling of hot carriers. Indeed, from the previous fit, we have obtained that the exciton formation has a lifetime for the lowest fluence ($\tau_{ef} = 1/(k_{ef} * a)$) of 120 fs (9 fs for the highest), while $\tau_c$ is 190 fs. Therefore, a lengthening of this decay can be associated with a longer cooling. As opposed to the PEA$_2$PbI$_4$ case (Figure 3C), the data cannot be fitted with only one $k_c$ shared among all traces (Figure S8, left). Thus, we proceeded to fit both the PEA$_2$PbI$_4$ and BuA$_2$PbI$_4$ data with separate $k_c$ values for each trace. The results are shown in Figures 4 and S8. While PEA$_2$PbI$_4$ does not show a clear trend, in BuA$_2$PbI$_4$, the cooling time clearly increases with fluence. This could be caused by a phonon bottleneck effect similar to that found in bulk lead halide perovskites.[46,55,56] Interestingly, an enhanced phonon bottleneck was observed for quasi-2D (n>1) perovskites containing BuA.[45] On the other hand, nanocrystals tend to show lengthened cooling, even allowing the extraction of hot carriers.[19] Recently, it was shown that PEA$_2$PbI$_4$ presents faster cooling kinetics (220 fs, very similar to our value of 190 fs) than 2D perovskites containing large permittivity cations (HOC$_2$H$_4$NH$_2^+$, 720 fs, and HOC$_3$H$_6$NH$_2^+$, 640 fs).[18] Although bottleneck effects were not explored, the difference was explained, with the help of *ab initio* simulations, as a combination of 1) a difference in screening due to the permittivity of the cations; 2) different nonadiabatic couplings between conduction bands; 3) different electron-phonon couplings; and 4) suppressed rotation of the spacers. While the first factor is not applicable to our cations, the rest may produce the difference in cooling we observe. Future work should address this question. Lastly, we were unable to obtain 2D TRTS maps or FLUPS spectra of BuA$_2$PbI$_4$. We did find that this 2D perovskite was much less stable under illumination, especially in the presence of air. We show the effects of this degradation on the recorded FLUPS spectra in Figure S9. Nonetheless, the BuA$_2$PbI$_4$ sample was stable for long enough to obtain the dynamics shown in Figure 4, owing to the inert atmosphere used in TRTS. However, 2D maps require a considerable fluence for more than 24 h, which produced a certain degree of degradation. In summary, for applications where both stability and fast carrier cooling are critical, PEA$_2$PbI$_4$ is the better option.

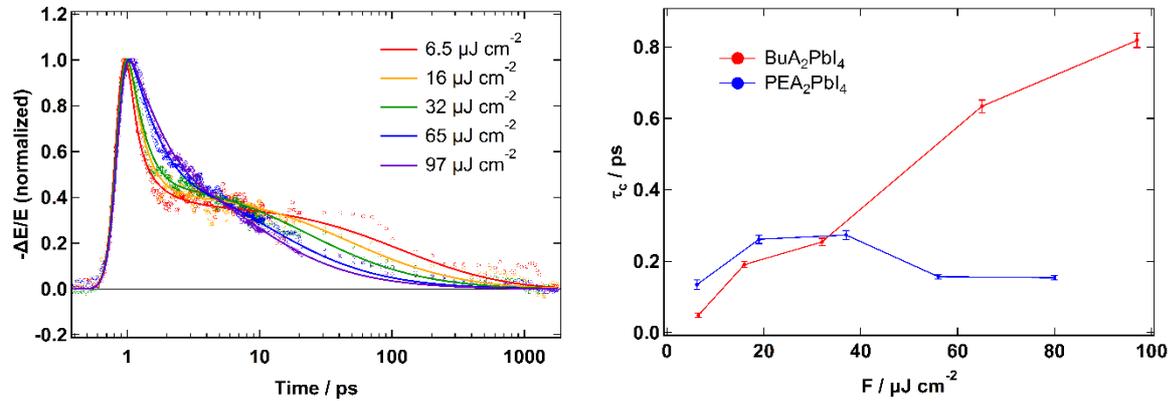

**Figure 4.** Left) TRTS decay dynamics of BuA$_2$PbI$_4$ with the global fit results obtained using independent $k_c$ values for each trace. Right) $\tau_c$ ($1/k_c$) values for the two different compositions at different fluences.

We have studied in detail the evolution of carriers and excitons in 2D perovskites using a combination of sensitive spectroscopic techniques. Indeed, we have shown the potential of using ultrabroadband time-resolved terahertz spectroscopy to study the evolution of carriers into excitons with a large binding energy at room temperature by following the spectral changes. In combination with fluorescence upconversion spectroscopy, we have demonstrated sequential cooling and exciton formation where exciton-exciton interactions play an important role. These are the source of a longer-lived population of carriers despite the large correlation induced by the quantum and dielectric confinements. Furthermore, a Mott transition, as the source of these carriers, has been discarded at these excitation densities through fluence dependent analysis of the fluorescence upconversion spectra. We have presented and applied a kinetic model that can explain our data through the use of global fits. Finally, we have used the previously presented work to explore the influence of the organic cation. We find that although the changes in the steady-state photophysics are small, the choice of the cation is crucial for both stability and hot carrier cooling, strongly affecting phonon bottleneck effects, which is a phenomenon to be considered for future applications.

Methods

Sample preparation – Similar to the bulk perovskites in our previous study,[26] thin film samples were prepared by spin-coating using an antisolvent method[57] on fully transparent HDPE substrates (1 × 15 × 15 mm). The substrates were cleaned in a Hellmanex® solution using a sonicating bath and subsequently rinsed with ultrapure water, acetone, ethanol and high purity methanol. Directly afterward, the substrates were submitted to an oxidative plasma treatment for 90 min. This treatment allowed better wettability of the apolar polymer with polar solutions through the creation of oxidized polar groups on the surface. Precursor solutions (BuAI or PEAI, 1 mmol, and $PbI_2$, 0.5 mmol, in 1 mL of DMSO) were prepared for deposition. Spin-coating was carried out as follows. A sufficient amount of solution was deposited over the substrate. The sample was spun at 1000 rpm for 10 s and subsequently at 6000 rpm for 30 s. Ten seconds before the end, 200 μL of chlorobenzene was poured onto the spinning substrate to act as an antisolvent and improve crystallization. Finally, the sample was annealed at 100°C to obtain a perovskite layer of the characteristic color.

Ultrabroadband time-resolved THz spectroscopy – TRTS measurements were carried out with a previously described laser spectrometer.[26,58] Three beams were split from the fundamental output (45 fs pulse duration, $\lambda$ = 800 nm, 1 kHz repetition rate) of an amplified CPA Ti:sapphire laser (Libra USP-Model, Coherent) and used for the TRTS experiments. The first beam was employed to pump a white-light-seeded optical parametric amplifier (OPerA Solo, Coherent) that provided the pump pulses for the pump–probe experiment at tunable wavelengths. Alternatively, the same beam could be diverted to obtain 400 nm pulses through second harmonic generation in a BBO crystal. The more powerful of the two remaining beams (390 μJ per pulse) was used to generate the probe beam, consisting of a train of short and broadband THz pulses, through a two-color plasma method:[59] The beam was focused with a fused silica lens ($f$ = 75 mm), and the second harmonic was generated with a 100 μm-thick BBO crystal. At the focal point, the electric field of the two-color beam was sufficiently strong to form a plasma filament in nitrogen that radiated a broadband THz pulse (200 fs, 1–20 THz) that was subsequently collimated and focused with parabolic gold mirrors onto the sample. To maximize THz generation, a dual wavelength waveplate was placed immediately after the BBO crystal to obtain fundamental and second harmonic beams of equal polarization after type 1 phase matching.[60] The transmitted beam went through two additional parabolic mirrors towards a homemade ABCD (air-biased coherent detection) detector.[61] Silicon wafers were used to filter the visible light from the THz generation and pump light. The remaining beam (40 µJ/pulse) was used as a gate for detection, generating a second harmonic signal proportional to the THz electric field measured with a PMT (PMM01, Thorlabs). The SHG process was carried out in an enclosed box with TPX® windows, where the atmosphere could be replaced with pure butane gas.[62] This setup allowed an increase in the sensitivity at the expense of frequencies in the 12-20 THz range. It was used to increase the signal-to-noise ratio in the frequency-averaged dynamics, where the $-\Delta E/E$ signal was obtained following the maximum signal point. The pulse shape and the effect of butane can be found in our previous study.[26]

Broadband fluorescence upconversion spectroscopy – Time-resolved emission measurements were achieved using the same CPA Ti:sapphire laser as for TRTS. Seventy-five percent of the beam was transmitted to a white-light-seeded optical parametric amplifier (OPerA-Solo, Coherent) to generate 110 µJ gate pulses at $\lambda$ = 1300 nm. The rest of the original beam was frequency doubled to 400 nm by a type I BBO crystal and set to the magic angle (54.7°) to consider only the population dynamics. A small beam stop and a 400 nm filter was placed after the sample position to block most of the transmitted 400 nm pump light. The horizontally polarized gate beam and vertically polarized fluorescence interacted in the 100 µm thick BBO crystal (EKSMA Optics), which had an optical axis in the horizontal plane. The upconverted signal was generated by type II sum frequency generation since the two inputs had different polarizations. This configuration is suitable for obtaining a broad frequency range without any moving part. The large angle between the fluorescence and the gate beam, here 21°, helped the phase matching requirement and the background-free detection of the signal. The signal was focused onto a fiber by a concave mirror while the frequency-doubled gate beam and the upconverted pump beam were sent away. The fiber transmitted the light to an unfolded Czerny-Turner spectrograph (Triax 320, Jobin Yvon). The incoming signal was separated by wavelength through a UV grating and sent to a CCD detector (Newton 920, Andor). The dynamics of the fluorescence signal were obtained with a computer-controlled delay stage (PI) in the pump path. The time correction for the instrument response function (IRF) was calculated as 190 fs using the cross correlation between the pump and the probe.

## Acknowledgement

Financial support by the Swiss National Science Foundation (SNSF, Grant No. 200021_175729) and the National Center of Competence in Research "Molecular Ultrafast Science and Technology" (NCCR-MUST), a research instrument of the SNSF, is gratefully acknowledged. The authors also thank Linfeng Pan for his help with SEM measurements.

## Supplementary Information

Additions to Fig. 1 – 2D maps of the real and imaginary parts of the THz signal reported in Fig. 2 – Detailed analysis of emission spectra measured at increasing energy fluences – Details of the TRTS data treatment – Details of the kinetic models and of the global fit – Supplementary experimental data: a) Influence of the cation upon the TRTS dynamics, and b) Stability of the samples as probed by FLUPS – Spectroscopic and SEM characterization of the samples.

# Exciton and Carrier Dynamics in 2D Perovskites

Andrés Burgos-Caminal, Etienne Socie, Marine E. F. Bouduban, and Jacques-E. Moser *

Photochemical Dynamics Group, Institute of Chemical Sciences and Engineering,
and Lausanne Centre for Ultrafast Science (LACUS),
École Polytechnique Fédérale de Lausanne, 1015 Lausanne, Switzerland

* Corresponding author, email: je.moser@epfl.ch

# Supporting Information

## S1. Main results

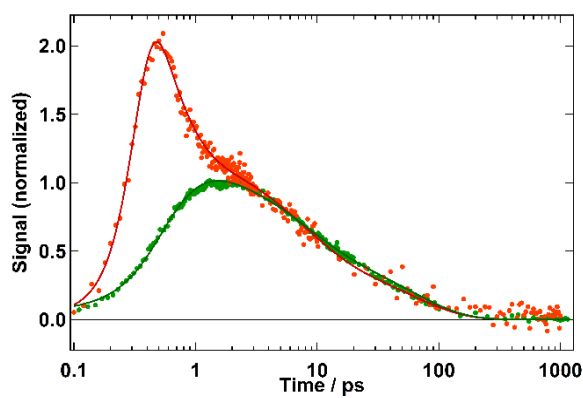

**Figure S1.** Full view of the decay in Figure 1A with a different normalization to highlight the coinciding decays.

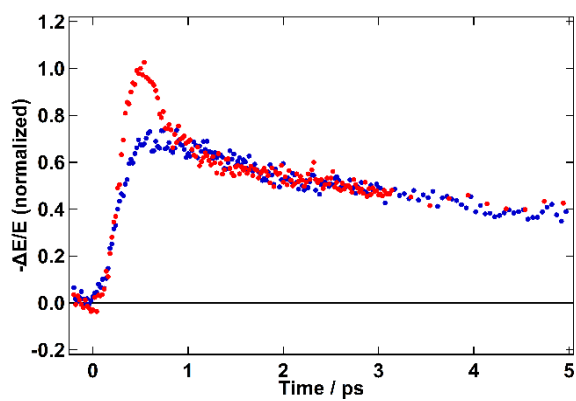

**Figure S2.** Figure 1B with a different normalization to highlight that the main difference lies in the initial decay.

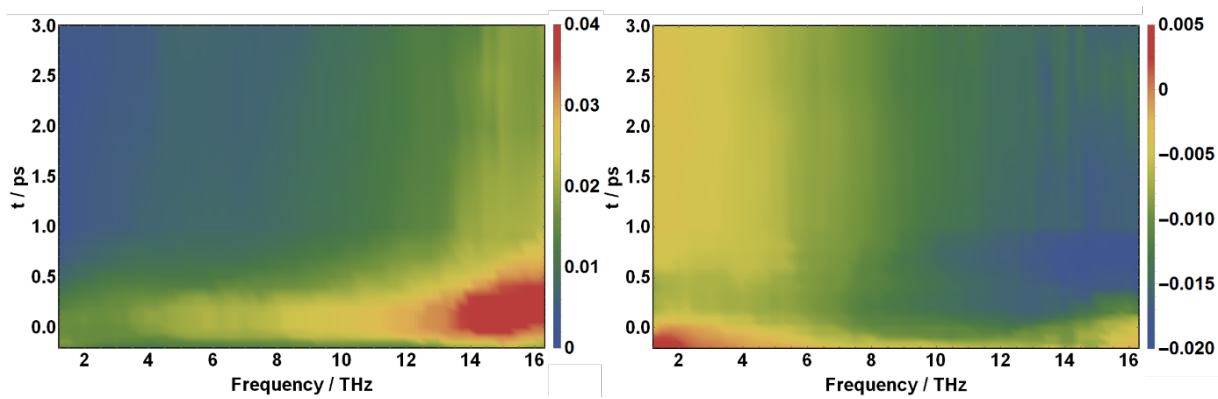

**Figure S3.** 2D maps of the real (left) and imaginary (right) parts of the −ΔE/E signal corresponding to the data in Figure 2. The maps have been cleaned with a median filter.

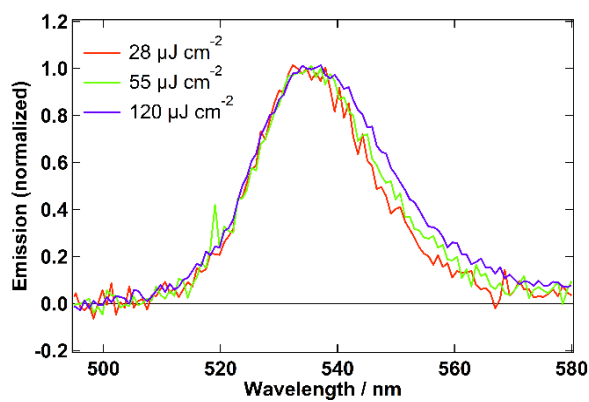

**Figure S4.** Normalized emission obtained from averaging the 0.96-1.04 ps range for different fluences. At higher fluence, additional contributions are only seen on the low energy side of the band.

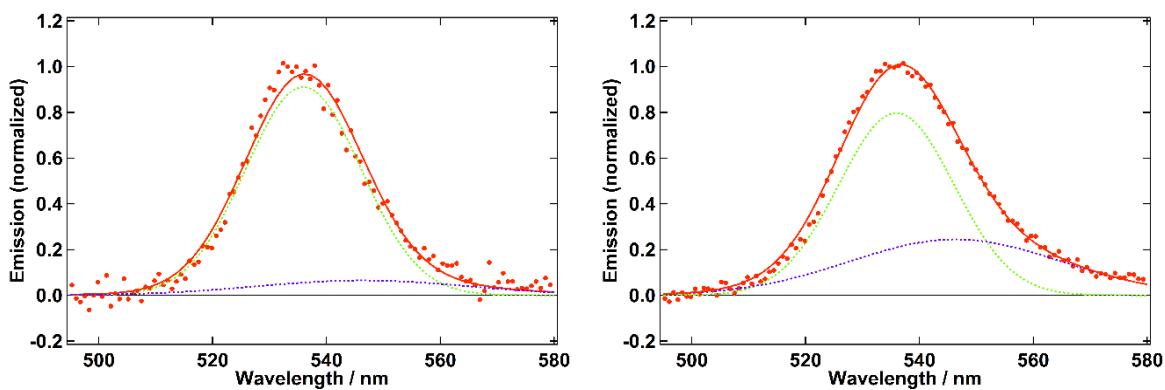

**Figure S5.** Multipeak fit of the emission at 1 ps for 28 μJ cm$^{-2}$ (left) and 120 μJ cm$^{-2}$ (right). The main peak is centered at 536 nm while the second peak is the result of a red shift of 44 meV. The second peak is clearly more important at a higher fluence. The widths and positions are equal for the two fits.

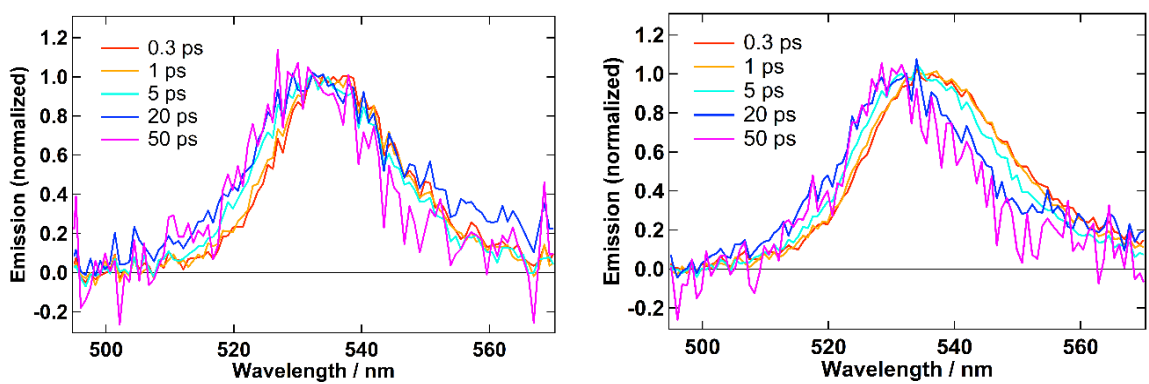

**Figure S6.** Normalized emission at 28 μJ cm$^{-2}$ (left) and 120 μJ cm$^{-2}$ (right) for different time delays. There is an appreciable blue shift that is not fluence dependent. The biexcitonic contribution on the red side is clearly dependent on the fluence and follows different kinetics.

**Table S1.** Fluence dependence of the tri-exponential fit for the $-\Delta E/E$ decay at $\lambda_{ex}$ = 400 nm. The first time constant can be related to carrier cooling while the latter two can be assigned to recombination.

| $F$ / μJ cm$^{-2}$ | $\tau_1$ / ps | $\tau_2$ / ps | $\tau_3$ / ps |
|---|---|---|---|
| 6.2 | 0.29 ± 0.03 | 12.5 ± 2.2 | 360 ± 35 |
| 19 | 0.29 ± 0.02 | 9.8 ± 1.1 | 248 ± 36 |
| 37 | 0.31 ± 0.02 | 8.5 ± 0.8 | 198 ± 35 |
| 56 | 0.29 ± 0.02 | 7.1 ± 0.6 | 111 ± 21 |
| 80 | 0.33 ± 0.02 | 6.1 ± 0.5 | 92 ± 16 |

## S2. TRTS data treatment

In order to obtain the data reported in Figure 2A, we need to calculate $\Delta\sigma(\omega)$. For that, we can use the following equation[1]

$$\Delta\sigma(\omega) = -\frac{\varepsilon_0 c (n_A + n_B)}{L} \frac{\Delta E(\omega)}{E(\omega)} \ ,$$

where $n_A$ = 1.00 and $n_B$ = 1.54 are the refractive index of the air and of the HDPE substrate, $\varepsilon_0$ is the vacuum permittivity, $c$ is the speed of light, $L$ = 550 nm is the thickness of the photoactive layer (Figure S11), and $\Delta E$ and $E$ are the Fourier transforms of difference in THz electric field upon photoexcitation and the original one, respectively. In the text, we often refer to $-\Delta E/E$ as the THz absorption signal. This equation is valid for thin films assuming that $\Delta E$ is small compared to $E$. In addition, this equation is valid under a steady-state condition, where $\Delta\sigma(\omega)$ does not substantially change during the duration of the pulse. However, this is not always the case, especially at very short pump-probe delays. A solution was introduced by Beard et al. where a 2D map is obtained and $\Delta\sigma(\omega, \tau)$ is obtained along the diagonal.[2] Likewise, $\Delta E$ can directly be measured scanning simultaneously the pump-probe and the gate-probe delays, allowing to obtain $\Delta E$ waveforms where all the data points have a constant delay ($\tau$) from the arrival of the pump.[3]

## S3. Kinetic model

The kinetic model presented in this work is based on following the densities of excitons ($N_{ex}$), hot carriers ($N_{hc}$), and cold carriers ($N_{cc}$) over time ($t$). Thus, we define three rate equations:

$$\frac{dN_{ex}(t)}{dt} = a\big(G(t)\big)_{510\ nm} + k_{ef}N_{cc}(t)^2 - k_{er}N_{ex}(t) - k_a N_{ex}(t)^2 - k_{ed}N_{ex}(t),$$

$$\frac{dN_{hc}(t)}{dt} = a\big(G(t)\big)_{400\ nm} + k_a N_{ex}(t)^2 - k_c N_{hc}(t),$$

$$\frac{dN_{cc}(t)}{dt} = -2k_{ef}N_{cc}(t)^2 + k_c N_{hc}(t) + 2k_{ed}N_{ex}(t)\ ,$$

where $k_{ef}$, $k_{er}$, $k_a$, $k_{ed}$, and $k_c$ are the kinetic constants for exciton formation, exciton recombination, exciton-exciton annihilation, exciton dissociation, and carrier cooling, respectively. In addition, $G(t)$ is the generation term defined as:

$$G(t) = \sqrt{\frac{4\log(2)}{\pi\ w^2}}\exp\left(-\frac{4\log(2)\ (t-t_0)^2}{w^2}\right),$$

where $w$ is the fwhm of the Gaussian. A term "2" is used to indicate that two carriers are taken or generated when one exciton is formed or dissociated. Alternatively, no such term is used for the Auger process since two excitons turn into two hot carriers. Furthermore, $k_{ed}$ is considered to be effectively zero, due to the large $E_b$ (270 meV). We can approximately consider that $k_{ed} = k_{ef}/\exp\left(-\frac{E_b}{k_b T}\right) = k_{ef}/4.4 \cdot 10^4 \approx 0$.

The signal is expressed as $S = N_{cc} + c_{hc}N_{hc} + c_{ex}N_{ex}$, where $c_{ex}$ and $c_{hc}$ are the proportionality coefficients for excitons and hot carriers, respectively. $a$ is the amplitude coefficient in density units. It serves as a measure of the fluence. During the global fit, only one value of $a$ is fitted for each wavelength. The different values for each fluence are obtained multiplying that value by a coefficient $b = F_i/F_0$ where $F_i$ is the fluence used in that particular trace and $F_0$ is the lowest fluence used.

The equations are solved numerically and fitted using a home-written code in *Wolfram Mathematica*. The results of the fit are shown in Table S2.

**Table S2.** Results of the global fit with their corresponding fit error, where "adu" stands for arbitrary density unit.

| Parameter | Value |
|---|---|
| $k_{er}$ / ps$^{-1}$ | 8.58×10$^{-4}$ ± 1×10$^{-5}$ |
| $k_{ef}$ / ps$^{-1}$ adu$^{-1}$ | 1.14 ± 2×10$^{-2}$ |
| $k_a$ / ps$^{-1}$ adu$^{-1}$ | 8.84×10$^{-3}$ ± 1×10$^{-4}$ |
| $k_c$ / ps$^{-1}$ | 5.26 ± 0.16 |
| $a_{400\,nm}$ / adu | 7.47 ± 0.22 |
| $a_{510\,nm}$ / adu | 9.02 ± 0.09 |
| $c_{ex}$ / - | 0.48 ± 0.02 |
| $c_{hc}$ / - | 0.57 ± 0.03 |

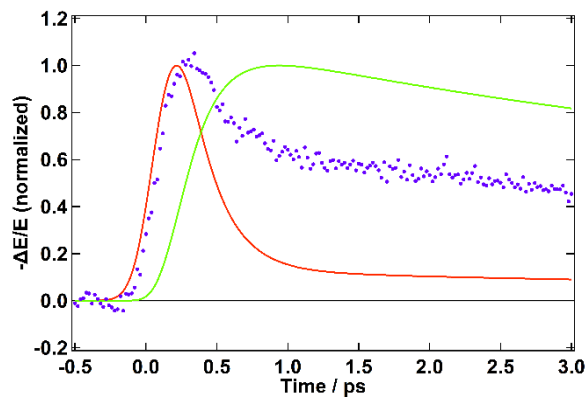

**Figure S7.** Exciton (green) and carrier (red) evolution according to the global fit on the kinetic model. Excitons show a delayed rise and a certain concentration of carriers is maintained due to Auger heating.

## S4. Influence of the cation

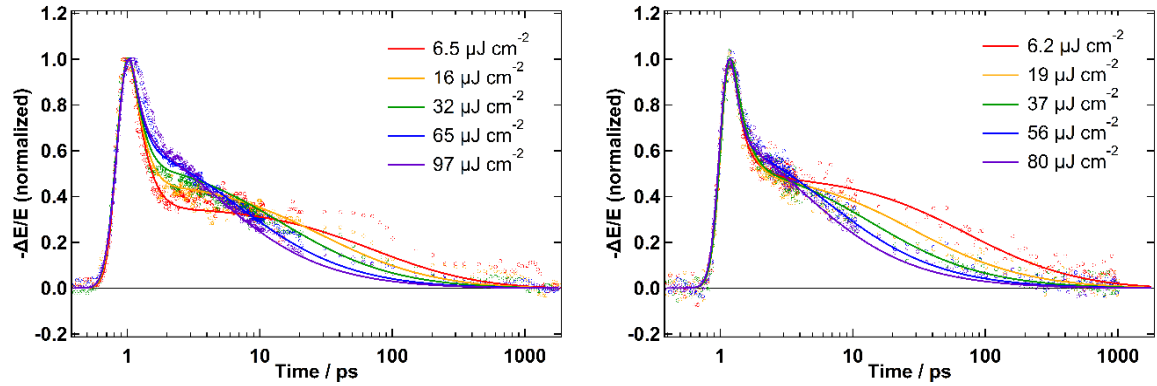

**Figure S8.** Left) BuA$_2$PbI$_4$ TRTS dynamics globally fitted using a single value for $k_c$. Clearly, there is a deviation between the fit and the data at early times. Right) PEA$_2$PbI$_4$ TRTS dynamics globally fitted using different values for $k_c$. The results do not vary substantially from the results with only one $k_c$, unlike BuA$_2$PbI$_4$ (Figure 4C).

**Table S3.** Results for the global fit with independent $k_c$ for each trace.

|  | BuA$_2$PbI$_4$ | PEA$_2$PbI$_4$ |
| --- | --- | --- |
| $k_{er}$ / ps$^{-1}$ | 1.22×10$^{-3}$ ± 2×10$^{-5}$ | 9.85×10$^{-4}$ ± 1.6×10$^{-5}$ |
| $k_{ef}$ / ps$^{-1}$ adu$^{-1}$ | 1.65 ± 7×10$^{-2}$ | 1.68 ± 5×10$^{-2}$ |
| $k_a$ / ps$^{-1}$ adu$^{-1}$ | 7.10×10$^{-3}$ ± 1.4×10$^{-4}$ | 7.57×10$^{-3}$ ± 1.3×10$^{-4}$ |
| $k_{c1}$ / ps$^{-1}$ | 20.5 ± 2.7 | 7.44 ± 7.5×10$^{-1}$ |
| $k_{c2}$ / ps$^{-1}$ | 5.24 ± 2.5×10$^{-1}$ | 3.82 ± 1.8×10$^{-1}$ |
| $k_{c3}$ / ps$^{-1}$ | 3.94 ± 1.6×10$^{-1}$ | 3.66 ± 1.7×10$^{-1}$ |
| $k_{c4}$ / ps$^{-1}$ | 1.58 ± 4×10$^{-2}$ | 6.37 ± 2.5×10$^{-1}$ |
| $k_{c5}$ / ps$^{-1}$ | 1.22 ± 3×10$^{-2}$ | 6.49 ± 2.7×10$^{-1}$ |
| $a_{400\,nm}$ / adu | 4.68 ± 6×10$^{-2}$ | 7.22 ± 1.4×10$^{-1}$ |
| $c_{ex}$ / - | 0.35 ± 1×10$^{-2}$ | 0.56 ± 3×10$^{-2}$ |
| $c_{hc}$ / - | 0.45 ± 1×10$^{-2}$ | 0.78 ± 4×10$^{-2}$ |

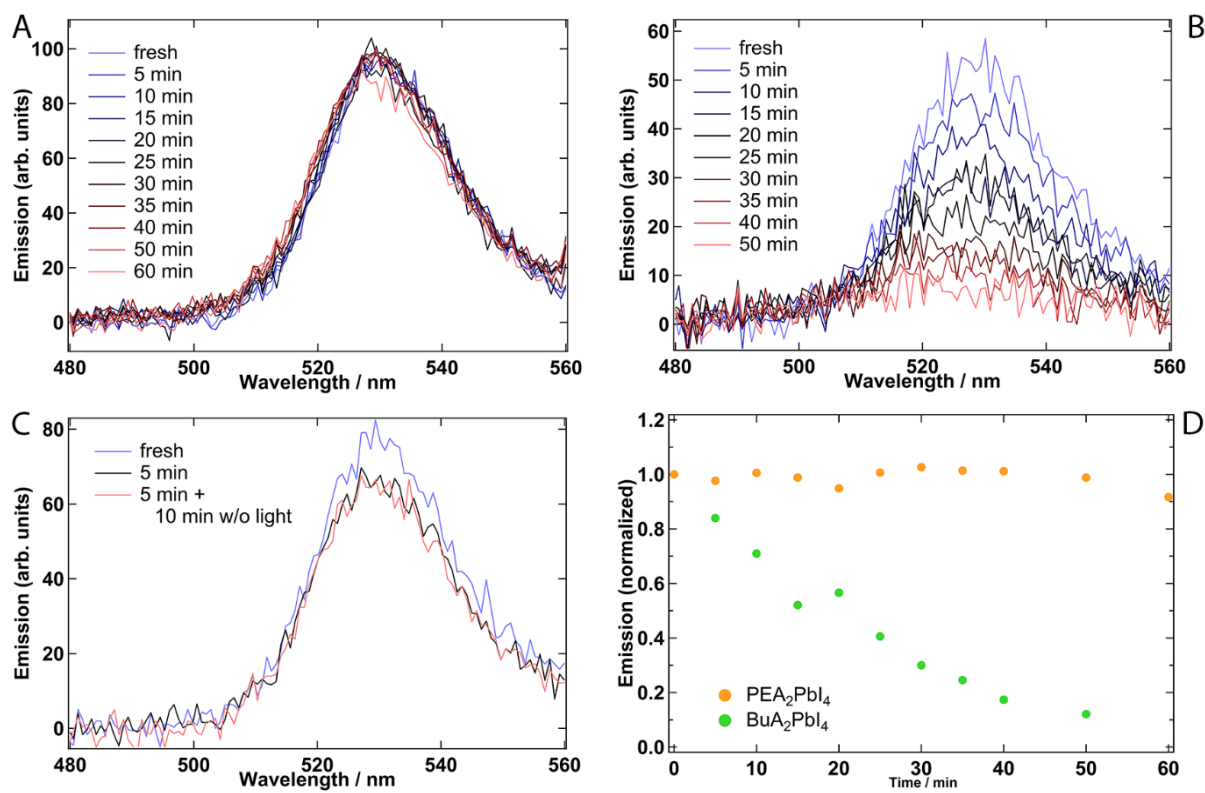

**Figure S9.** Stability comparison by following the FLUPS emission spectra between PEA$_2$PbI$_4$ (A) and BuA$_2$PbI$_4$ (B) under laser light illumination. Indeed, C shows that only under laser light the signal decays. In a few minutes the degradation is substantial for BuA$_2$PbI$_4$ rendering the FLUPS measurements impossible (D).

## S5. Sample Characterization

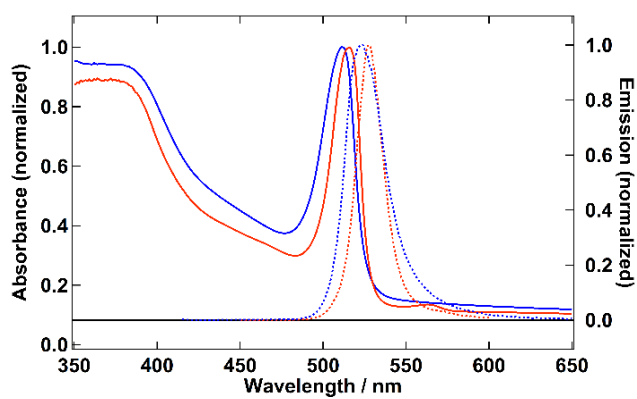

**Figure S10.** Absorption (solid line) and photoemission (dashed line) spectra of PEA$_2$PbI$_4$ (red) and BuA$_2$PbI$_4$ (blue). BuA$_2$PbI$_4$ shows a small blue-shift and broadening. The latter could be assigned to disorder.[4]

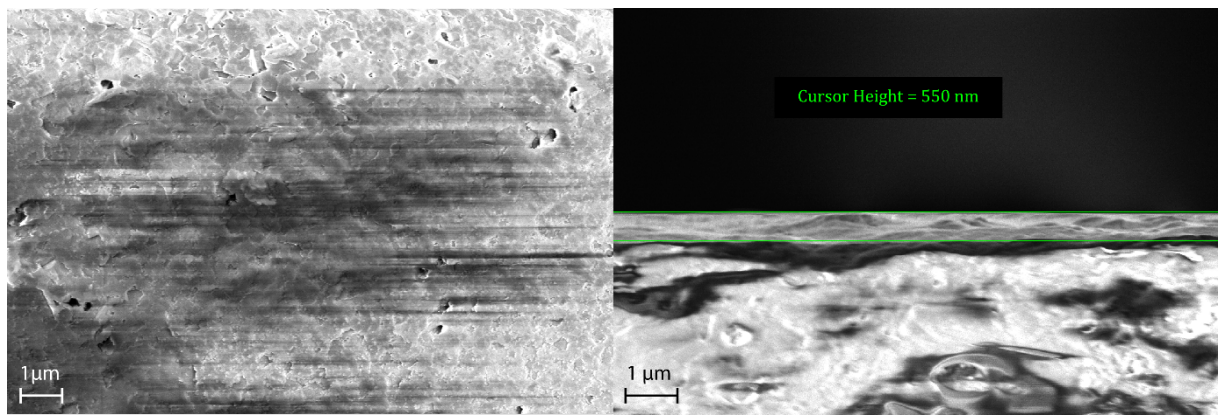

**Figure S11.** Left) SEM picture of PEA$_2$PbI$_4$. The flat microscopic crystals appear to lay horizontal to the surface. Right) Cross-sectional SEM showing a thickness of 550 nm for the PEA$_2$PbI$_4$ film.

## References of the SI